\documentclass[conference]{IEEEtran}
\usepackage[colorlinks,urlcolor=blue,linkcolor=blue,citecolor=blue]{hyperref}

\usepackage[hyphenbreaks]{breakurl}

\usepackage{color,array}

\usepackage{graphicx}
\begin{document}
\title{A Deep Dive into the Computational Fidelity of High Variability Low Energy Barrier Magnet Technology for Accelerating Optimization and Bayesian Problems
\thanks{\IEEEauthorrefmark{1}mm8by@virginia.edu\vspace{-1ex},\IEEEauthorrefmark{2}gangulys2@vcu.edu\vspace{-1ex}}}
\author{\fontsize{11}{11}\selectfont Md Golam Morshed\textsuperscript{1}\IEEEauthorrefmark{1}, Samiran Ganguly\textsuperscript{2}\IEEEauthorrefmark{2}, and Avik W. Ghosh\textsuperscript{1,3}\\
\fontsize{10}{12}\selectfont \textsuperscript{1}Department of Electrical and Computer Engineering, University of Virginia, Charlottesville, VA 22904, USA\\ \textsuperscript {2}Department of Electrical and Computer Engineering, Virginia Commonwealth University, Richmond, VA 23284, USA\\\textsuperscript {3}Department of Physics, University of Virginia, Charlottesville, VA 22904, USA
}
\maketitle
\begin{abstract}
Low energy barrier magnet (LBM) technology has recently been proposed as a candidate for accelerating algorithms based on energy minimization and probabilistic graphs because their physical characteristics have a one-to-one mapping onto the primitives of these algorithms. Many of these algorithms have a much higher tolerance for error compared to high-accuracy numerical computation. LBM, however, is a nascent technology, and devices show high sample-to-sample variability. In this work, we take a deep dive into the overall fidelity afforded by this technology in providing computational primitives for these algorithms. We show that while the compute results show finite deviations from zero variability devices, the margin of error is almost always certifiable to a certain percentage. This suggests that LBM technology could be a viable candidate as an accelerator for popular emerging paradigms of computing.
\end{abstract}


\section{INTRODUCTION}\label{introduction}

\begin{figure*}[!htbp]
    \centering
    \includegraphics[width=0.90\textwidth]{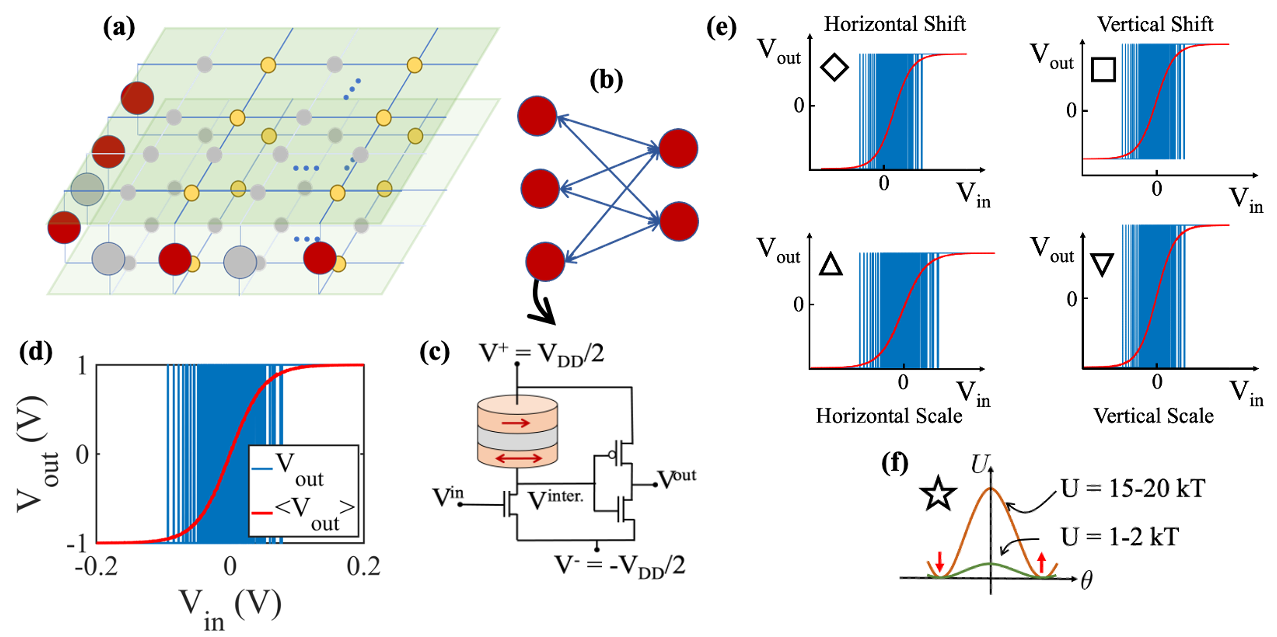}
    \caption{(a) Illustrative schematic of an embedded RBM, an energy-based optimization and learning algorithm, in a dual-stacked feedback cross-bar structure with neurons (the compute units) at the edges (large circles), while the synaptic weights (the program) loaded in
memristors located at the cross-points of the core cross-bar
structure (small circles). The active neurons and synapses are colored bold (red and yellow), while inactive units are greyed out. (b) The RBM network that gets embedded in (a). The bidirectional blue lines represent the synaptic connections between the neurons (red circles). The yellow circles used in (a) are not shown here for simplicity. (c) The design of an LBM-magnetic tunnel junction (MTJ)-based p-bit unit. (d) Ideal characteristics of a p-bit device. (e) Schematics of different characteristics distortions. (f) Illustration of energy barrier variation in a nanomagnet. Symbols (diamond, square, etc.) in (e) and (f) represent different variabilities henceforth.} 
    \label{fig1}
\end{figure*}

Low energy barrier magnet (LBM) technology, which utilizes nanomagnets with barrier height in the order of thermal energy, has recently been proposed as a potential candidate for hardware accelerators for probabilistic computing and stochastic sampling~\cite{bib1,bib2}. These accelerators may be broadly considered as hardware Markov chain Monte Carlo implementation that utilizes the built-in stochasticity provided by the dynamics of the LBM, which results in highly compact devices with true stochasticity, as compared to linear feedback-shift register (LFSR) based pseudo-random number generators (pRNGs)~\cite{bib3}. The magnetization component $m_z$ of the LBM randomly fluctuates between two stable states ($\uparrow,~ \downarrow$) under the influence of the thermal noise, and the probability of getting any one of the two stable states can be driven via an external current~\cite{bib4}. There are a handful of applications ranging from probabilistic computing to machine learning and artificial intelligence that leverage the intrinsic stochastic nature of LBMs~\cite{bib4,bib5,bib6,bib7,bib8}. The prototype hardware building blocks are the binary stochastic neurons (BSNs), popularly known as ``p-bits'' with programmable weights in a recurrent configuration. An illustrative example of a dual-stacked feedback cross-bar structure is shown in Fig.~\ref{fig1}(a). The synaptic weights or the ``program'' is loaded in memristors located at the cross-points of the core cross-bar structure, whereas the neurons are at the peripheries. Using a dual cross-bar structure, it is possible to build recurrent networks, including a Restricted Boltzmann Machine [RBM, Fig.~\ref{fig1}(b)], an example application area of this accelerator. The RBM is embedded in the computing fabric by enabling certain neurons and synaptic connections while disabling the rest. 

Although BSN-based non-Boolean probabilistic applications are inherently more error resilient than conventional nanomagnet switches used for deterministic Boolean memory and logic applications, the computational reliability of these accelerators that employ LBMs as their hardware RNG, needs to be carefully assessed. Recently, several studies have discussed the impact of geometric, structural, and process variation from device-to-device that can create ignorable to high variability in the characteristics of LBMs depending on the degree of variation~\cite{bib9,bib10,bib11}, however, the resulting impact of these ``non-idealities'' on the computational networks is still largely not understood.
%%%%%%%%%%%%%%%%%%%%%%%%%%%%%%%%%%

In this letter, we discuss the issues of variability in the context of circuits and networks built from LBM-based BSN devices. We categorize the variability into a few broad classes, namely shifting and scaling of the device characteristics from the ideal as expected from the mathematical model, and the variability of the barrier heights for two broad classes of algorithms that can be solved using p-bits, such as energy minimization-based optimization algorithm (EMOA) and probabilistic graphical algorithm (PGA). EMOA includes problems such as Ising model and RBMs, which seek to define a problem in terms of a thermodynamically definable ``energy-landscape'' with the embedding of the desired optimal result in the ground/vacuum energy, while PGA includes Bayesian decision diagrams, which do not have an inherent notion of energy and thermodynamics. In terms of network connectivity (using the spectral theorem of linear systems)~\cite{bib12}, this implies that the EMOA networks have symmetric or undirected connections, resulting in eigenstates that are real-valued and reachable via real-space computation, whereas PGA networks are asymmetric or directed, resulting in non-real or complex eigenstates not reachable via real-space computation.

We estimate the mean absolute error (MAE) to quantify the performance deviation from the ideal devices. We find the MAE shows a sub-linear saturation for EMOA, while in the PGA, the error grows linearly to super-linearly. Moreover, the networks are found to be more prone to shifting variability than scaling. Additionally, for EMOA, larger networks are less affected by the variability, while for PGA, the trend is the opposite. Our findings may provide a potential path forward toward designing reliable LBM-based hardware accelerators.

\section{BUILDING `p-bits' USING LOW BARRIER MAGNETS}\label{building-p-bit-lbm}
Thin film magnets used in magnetic random access memory (MRAM) technology exhibit a double potential well corresponding to the two easy points [Fig.~\ref{fig1}(f)]. The height of the barrier determines the expected state retention time using the Arrhenius relation given by:
\begin{equation}
\tau = \tau_0 e^{U/k_BT}
\label{eq:magnet-state-retention}
\end{equation}
In the above equation, $U (= \mu_0 M_s H_k \Omega/2)$ is the energy barrier, where the symbols respectively stand for permeability of free space, saturation magnetization, magnetic anisotropy field strength, and volume. For a conventional storage class memory, $U$ is set to $40 - 60~k_B T$ for the free layer of an MTJ, which yields a decade-long state retention time $\tau$ depending on the $\tau_0$, the inverse of attempt frequency that ranges from $0.1 - 1~ns$~\cite{bib13}. However, if the magnet is ultra-scaled by reducing the volume $\Omega$ or its profile is made circular, which reduces the $H_k$ by removing the shape anisotropy, the retention time can be scaled down to near $\tau_0$~\cite{bib14}. In this case, the free layer's magnetization vector fluctuates between the two easy points under the influence of the thermal noise, which is able to ``kick'' the magnetization over the barrier with ease, at near $\mathrm{GHz}$ frequencies. MTJ structure allows this fluctuation to be translated into an equivalent fluctuation in the resistance of the device, which can be used for building useful devices that can harvest true randomness from the environment.

One such device is the ``p-bit'', which is a binary stochastic neuron with a compact model given by:
\begin{equation}
V^{out}_i = \rm{sgn}[\tanh(\beta V^{in}_i) + \alpha\cdot\rm{rnd}(-1,+1) ]V_{DD}/2
\label{eq:pbit-eqn}
\end{equation}
In this device, the output swings between $-V_{DD}/2$ to $V_{DD}/2$ corresponding to $-1$ and $+1$ state labels of $m_z$, however, the ratio of these states is controllable by an input signal, which imposes a $\tanh$-like probability distribution. rnd is a uniform random distribution. The parameters $\beta$ and $\alpha$ represent the transfer gain of the unit and the relative contribution of the stochasticity to the characteristics, respectively. For large scale correlated networks, $V^{in}$ can be represented as:
\begin{equation}
V^{in}_i = \kappa[h_i+\sum_j{J_{ij}{V^{out}_j}/{(V_{DD}/2})]}
\label{eq:pbit-input}
\end{equation}
where $j$ stands for the index over all input devices connected to the particular $i$-th device, $h$ is the bias vector, and $J$ is the synaptic matrix. Different functionalities correspond to different choices of $h$ and $J$. $\kappa$ is a coupling coefficient representing the inverse of the ``temperature'' of the system.
\section{SIMULATION METHOD}\label{simulation-method}
We implement the compact model of $p$-bit networks described by (\ref{eq:pbit-eqn}) and (\ref{eq:pbit-input}) in MATLAB according to the methodology discussed in Camsari~\cite{bib4}. The MATLAB model is a parameterized version of the compact modeling simulation performed in SPICE~\cite{bib4,bib15}. In MATLAB implementation, we use $\alpha=1$, $\beta=1$ (for ideal case), $\kappa=0.8$, and $V_{DD}=2~V$ throughout the calculation unless otherwise specified.

We use computational networks constructed from p-bits of varying sizes. For EMOA, we use AND gate and full-adder having $J$ matrices sized $3\times 3$ and $14\times 14$, respectively~\cite{bib4}. We construct an arbitrary symmetric $J$ matrix of $50\times 50$ for a large network. For PGA, we use Bayesian networks (BNs) constructed from $8$, $20$, and $50$ p-bits ($J$ matrices are asymmetric in these cases). For EMOA, MAE is computed by taking the summation of the absolute difference between the output probability distribution of ideal and non-ideal cases, normalized by the number of LBMs in the network. However, for PGA, we calculate the normalized MAE from the difference in the correlation matrix ($\sigma (i,j)=\frac{1}{T}\int_{0}^{T}V_i^{out}V_j^{out}\,dt $) between the ideal and non-ideal cases. For both algorithms, we use $T=10^{6}$ simulation steps to get to the $V^{out}$. If the sample generation time is $2~ns$, this is equivalent to $2~ms$ of compute time. The average and standard deviation of the MAE are calculated from $N=100$ simulations.
\section{RESULTS}\label{results}
LBM devices are hybrids of silicon CMOS, which is a highly mature technology, and spintronics/magnetics, which is a relatively new technology. While they have been successfully integrated into the context of high energy barrier storage class MRAM technology by several commercial vendors, its LBM variant comes with lithographic challenges that may require a long process of technological developments to perfect. These lithographic challenges mainly concern the quality of magnetic films and the precision control over their geometry. Abeed {[}2019a{]} studied the impact of geometrical irregularities such as dimples, holes, shape variance, etc. on the characteristic correlation times of LBMs and found that the distribution of correlation times can be large. These kinds of variations can have implications that are beyond the intrinsic behavior of the free-layer magnet of the MTJ itself.

In particular, two critical sets of variations are discussed next. Please note that these variations become relevant in the context of circuits and networks built from these devices.
%%%%%%%%%%%%%%%%%%%%%%%%%%%%%%%%%%
\begin{figure}[!htbp]
    \centering
    \includegraphics[width=0.49\linewidth]{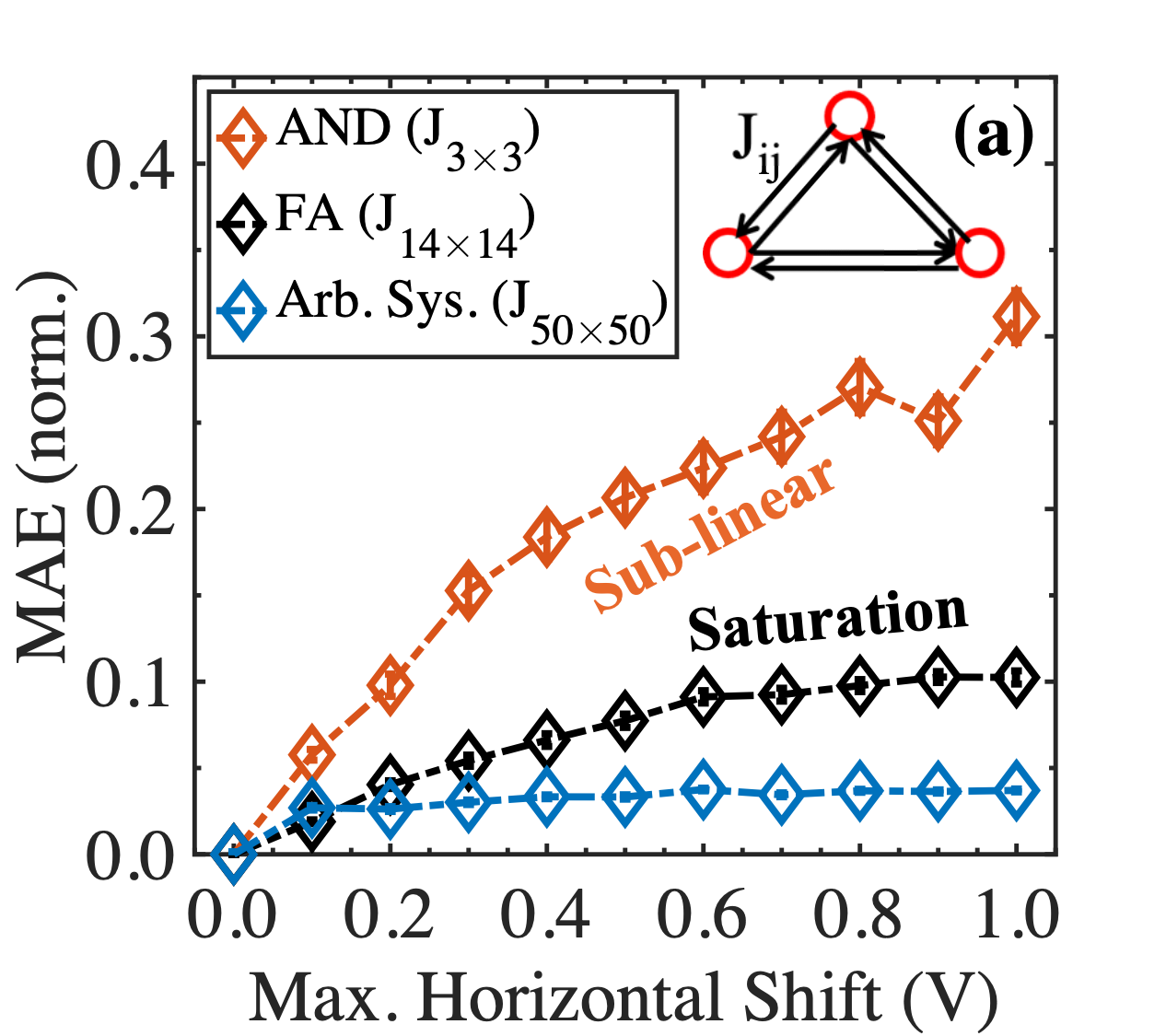}
    \includegraphics[width=0.49\linewidth]{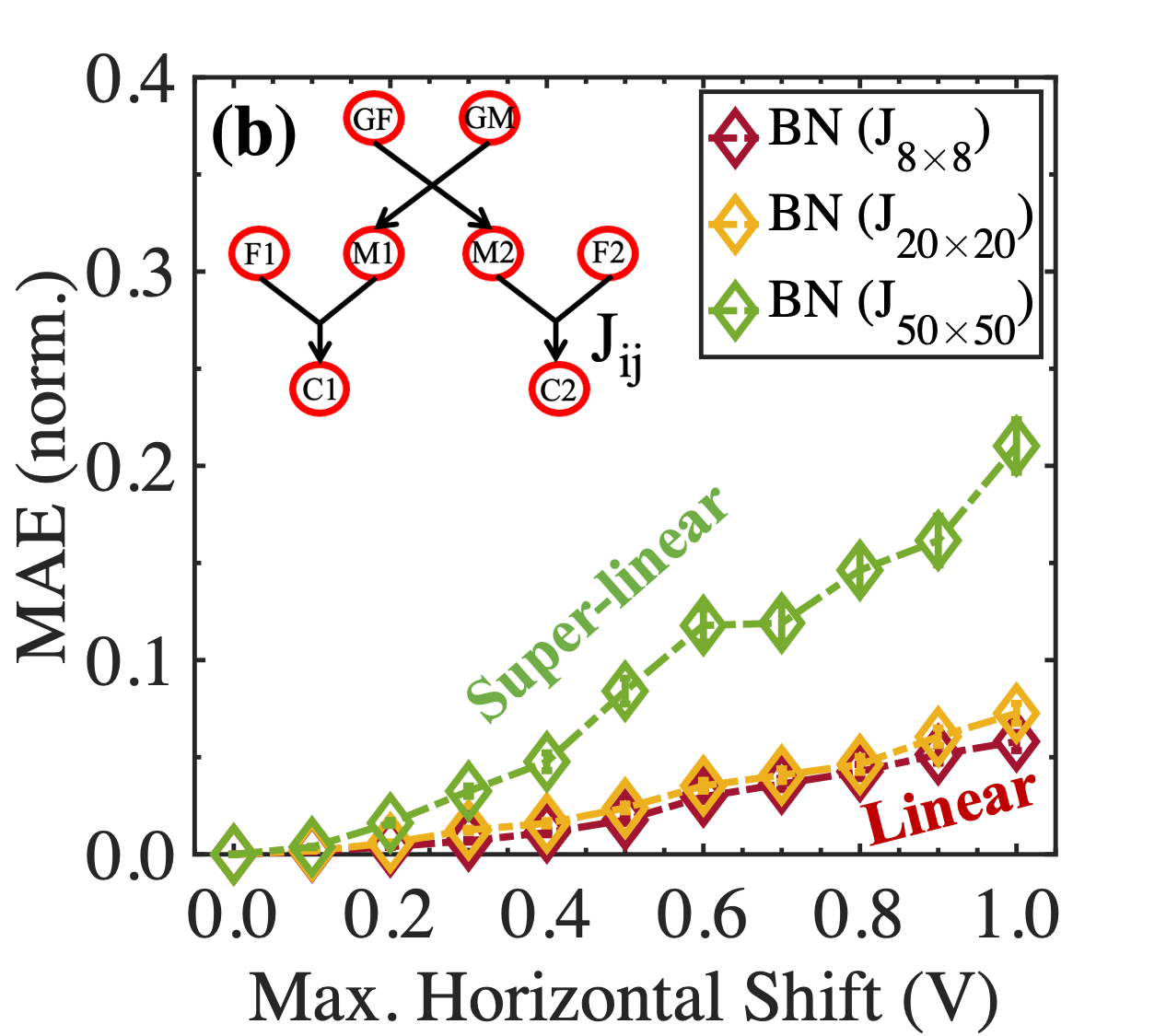}
    \caption{Normalized MAE from horizontal shifting for (a) EMOA and (b) PGA with different network sizes (size of the $J$ matrix). In all subsequent figures, we measured the average and standard deviation of the MAE from $100$ simulations. Colored lines in each figure represent different network sizes, and different symbols represent different distortions introduced in Fig.~\ref{fig1}(e). Inset in (a) shows the schematic of a $3 \times 3$ EMOA network, while inset in (b) shows the schematic of an $8 \times 8$ PGA Bayesian network representing a family tree (GF: Grandfather, etc.).} 
    \label{fig2}
\end{figure}
%%%%%%%%%%%%%%%%%%%%%%%%%%%%%%%%%%
\begin{figure}[!htbp]
    \centering
    \includegraphics[width=0.49\linewidth]{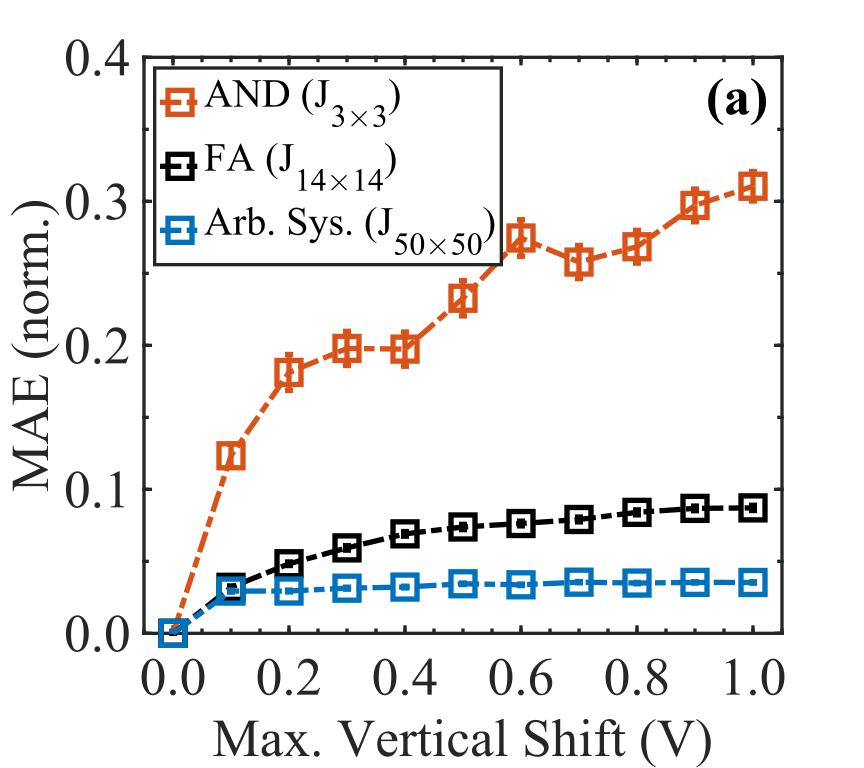}
    \includegraphics[width=0.49\linewidth]{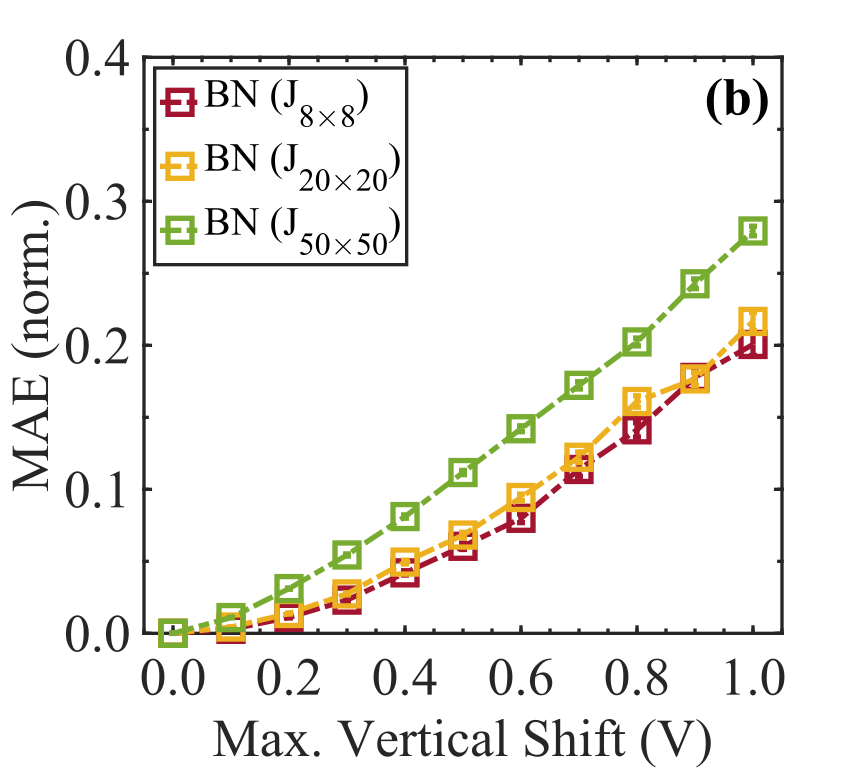}
    \caption{Normalized MAE from vertical shifting for (a) EMOA and (b) PGA with different network sizes.} 
    \label{fig3}
\end{figure}

%%%%%%%%%%%%%%%%%%%%%%%%%%%%%%%%%%
\begin{figure}[!htbp]
    \centering
    \includegraphics[width=0.49\linewidth]{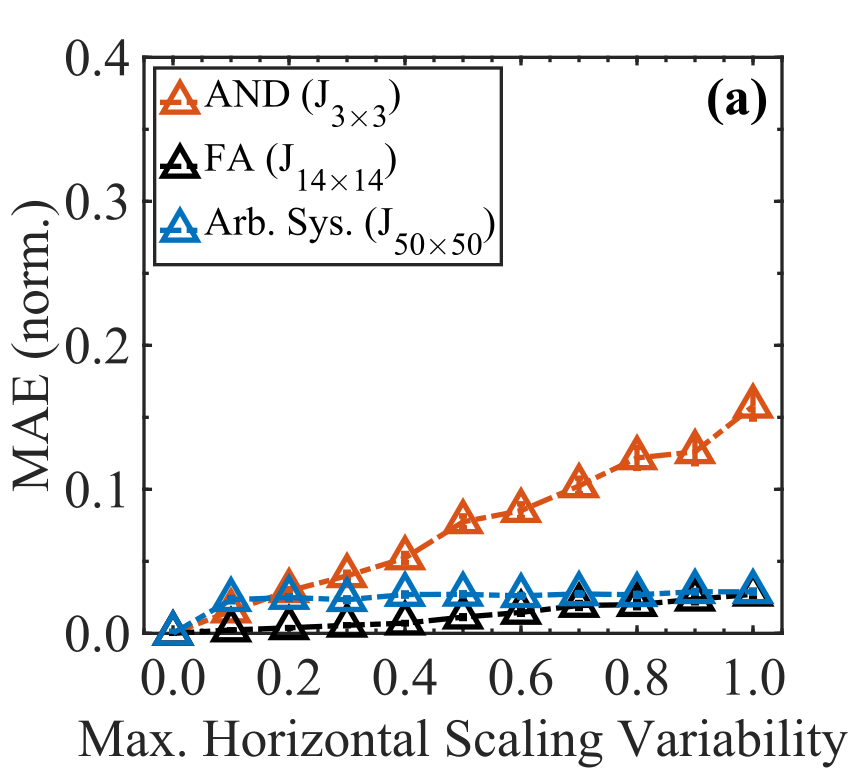}
    \includegraphics[width=0.49\linewidth]{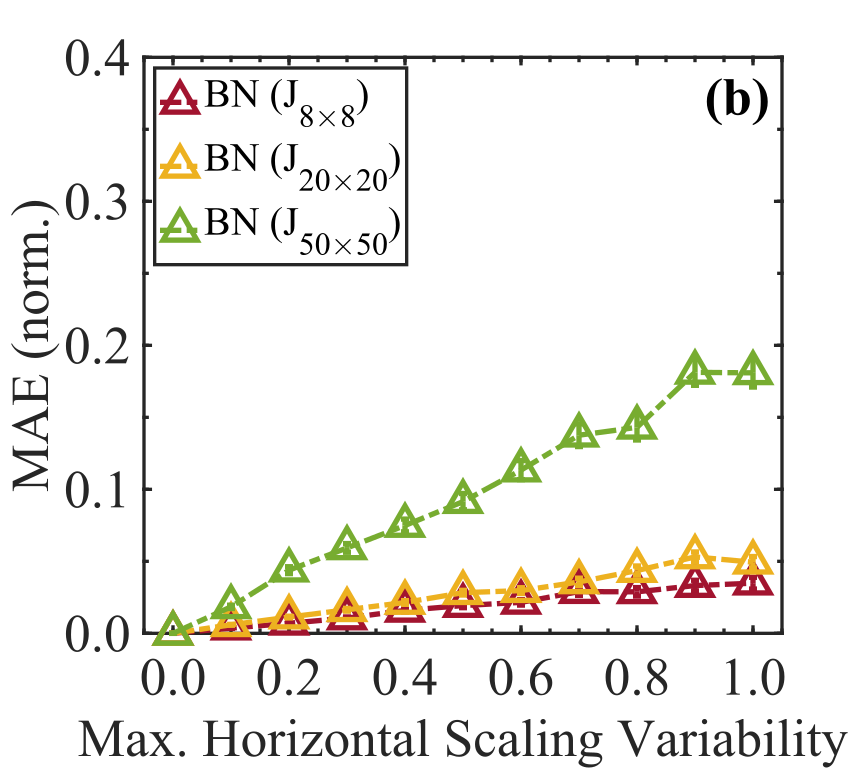}
    \caption{Normalized MAE from horizontal scaling for (a) EMOA and (b) PGA with different network sizes.} 
    \label{fig4}
\end{figure}
%%%%%%%%%%%%%%%%%%%%%%%%%%%%%%%%%%
\begin{figure}[!htbp]
    \centering
    \includegraphics[width=0.49\linewidth]{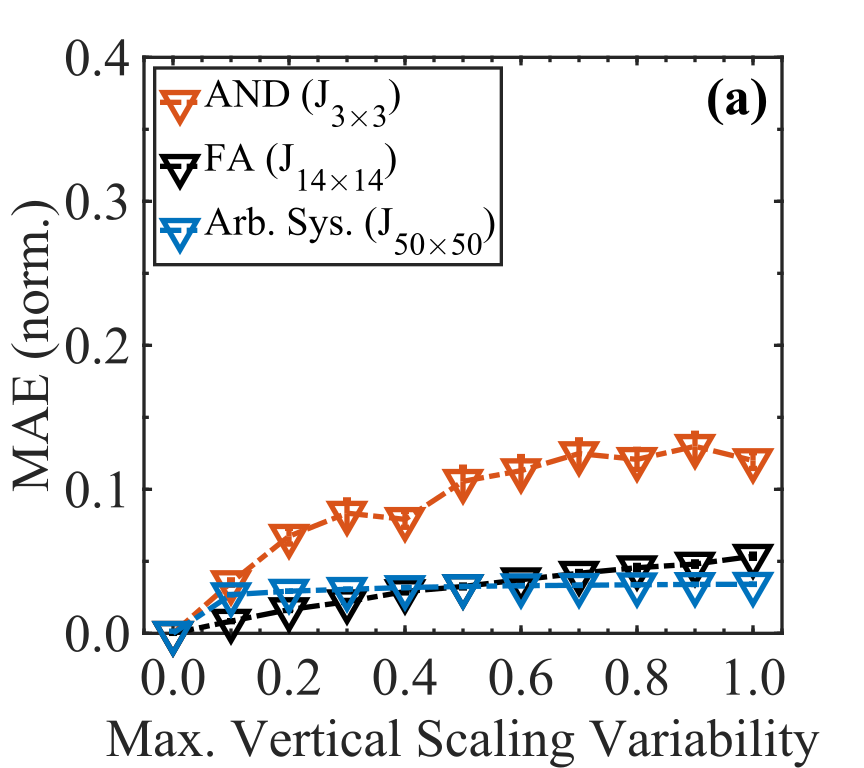}
    \includegraphics[width=0.49\linewidth]{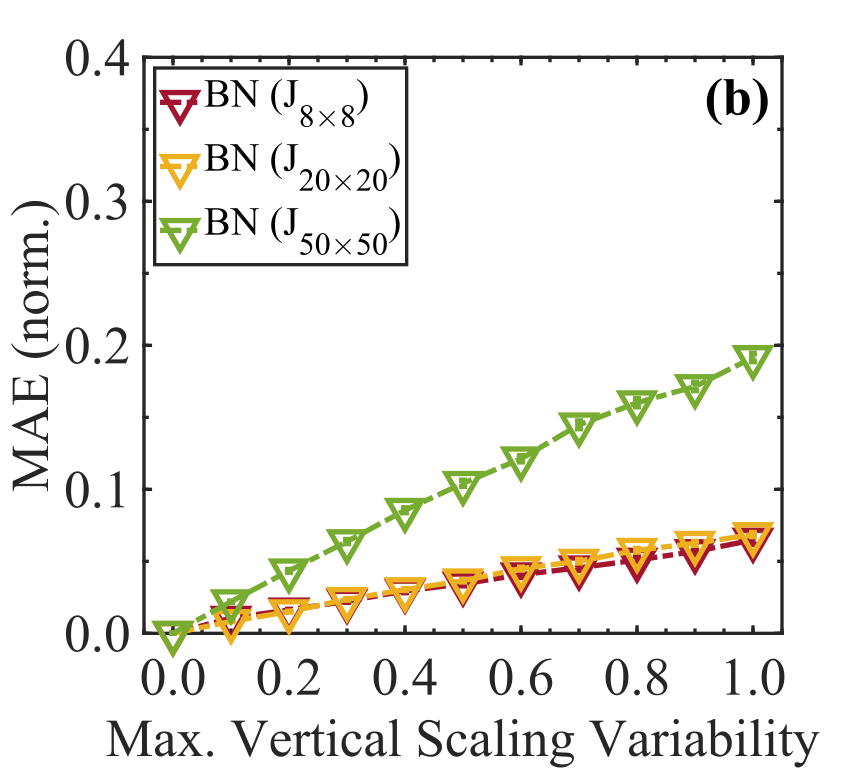}
    \caption{Normalized MAE from vertical scaling for (a) EMOA and (b) PGA with different network sizes.} 
    \label{fig5}
\end{figure}
\subsection{Characteristics Distortion}\label{title}
\begin{figure}[!htbp]
    \centering
    \includegraphics[width=0.98\linewidth]{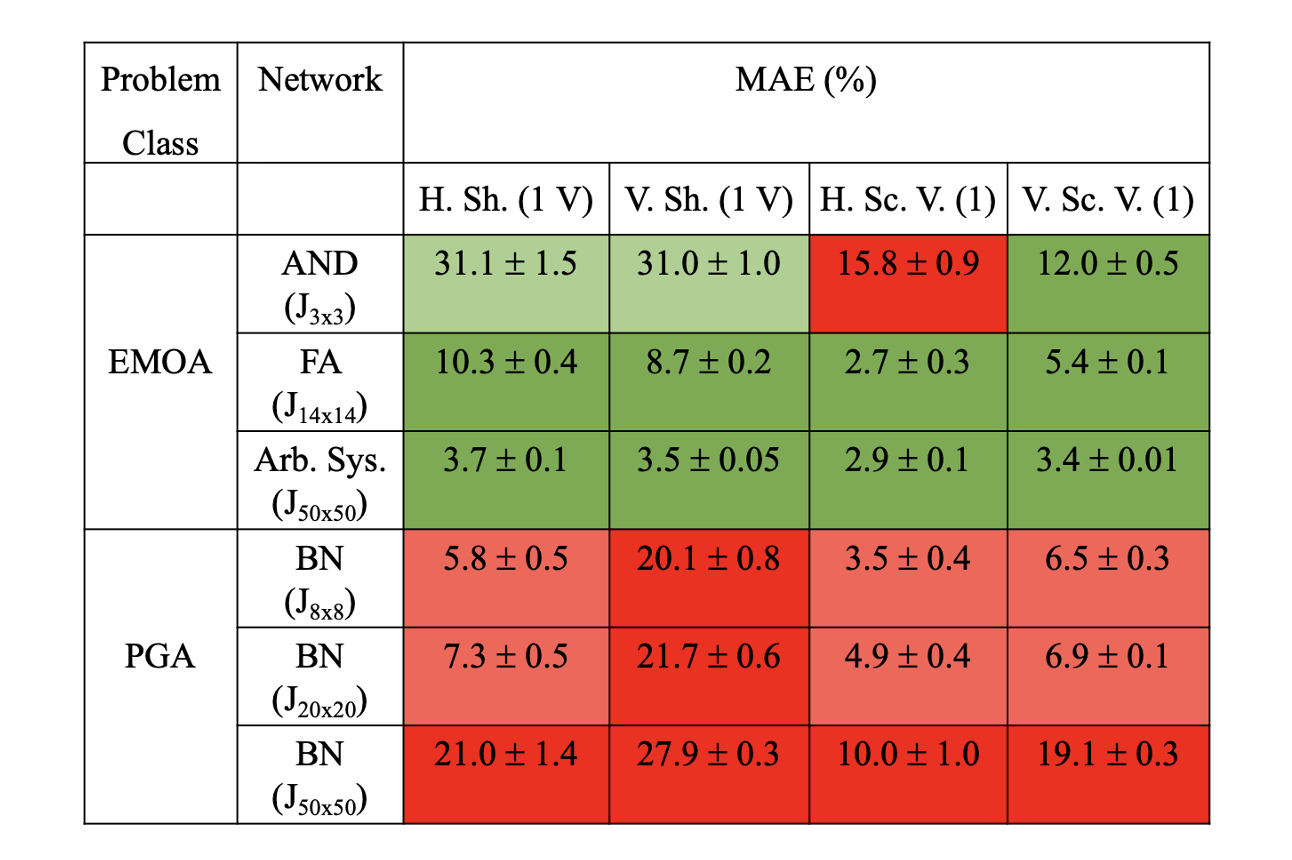}
    \caption{Normalized MAE arising from different characteristics distortions ($N=100$). The green, light green, light red, and red colors represent the saturation, sub-linear, linear, and super-linear trends of MAE, respectively, as a function of characteristics distortion.} 
    \label{fig6}
\end{figure}

Fig.~\ref{fig1}(c) and \ref{fig1}(d) show the proposed device and its ideal output characteristics, respectively. The characteristics of the device depend on the swing that is generated by the NMOS transistor turning on or turning off, balanced around the MTJ's characteristic resistance, i.e., the resistance of the transistor in the linear intermediate mode should match the MTJ's average resistance. In the linear mode of operation, the $\tanh$ shape shows up as an interplay between the MTJ's average and transistor's intermediate resistance as it swings from on to off, while the MTJ's magnetization flipping adds the fluctuation on the characteristics. A mismatch between these two can lead to a deviation from the ``ideal'' model presented in~(\ref{eq:pbit-eqn}). We categorize the variations into four categories that broadly cover the phase space of such distortions [shown in Fig.~\ref{fig1}(e)]: 1. horizontal shift; 2. vertical shift; 3. horizontal scale, due to variation in gain $\beta$; 4. vertical scale, from loading effects from follow on p-bits that the output cannot handle adequately. Fig.~\ref{fig2} shows the normalized MAE that emerged from horizontal shifting in the networks for both the EMOA and PGA problem classes. We vary the maximum voltage shift from $0~V$ to $1~V$. From Fig.~\ref{fig2}(a), we find that for AND gate, the error increases rapidly up to a $\sim 20\%$ horizontal voltage shift and slows down afterward. However, the error has an overall increasing trend. For larger networks, the error starts saturating at $\sim 10\%$ voltage shift. For AND gate, we find a maximum of $\sim 30\%$ MAE corresponding to a horizontal voltage shift of $1~V$. We see that as the network size increases, the error percentage decreases for EMOA. On the other hand, for PGA, from Fig.~\ref{fig2}(b), we can see that the error increases almost linearly as a function of horizontal voltage shift. However, the relation between the error and the network size is opposite to that of EMOA. Fig.~\ref{fig3}-\ref{fig5} show the normalized MAE that emerged from vertical shifting, horizontal scaling, and vertical scaling, respectively, for both EMOA and PGA. The increasing trend of the MAE is similar for different types of distortion; however, the percentage error varies depending on the problem class, distortion type, and network size. We list the maximum error arising from different types of distortion in Fig.~\ref{fig6}, where different colors represent the overall trend of MAE. It is important to note that we vary only one type of distortion at a time.
\subsection{Energy Barrier Variability}
\begin{figure}[!htbp]
    \centering
    \includegraphics[width=0.49\linewidth]{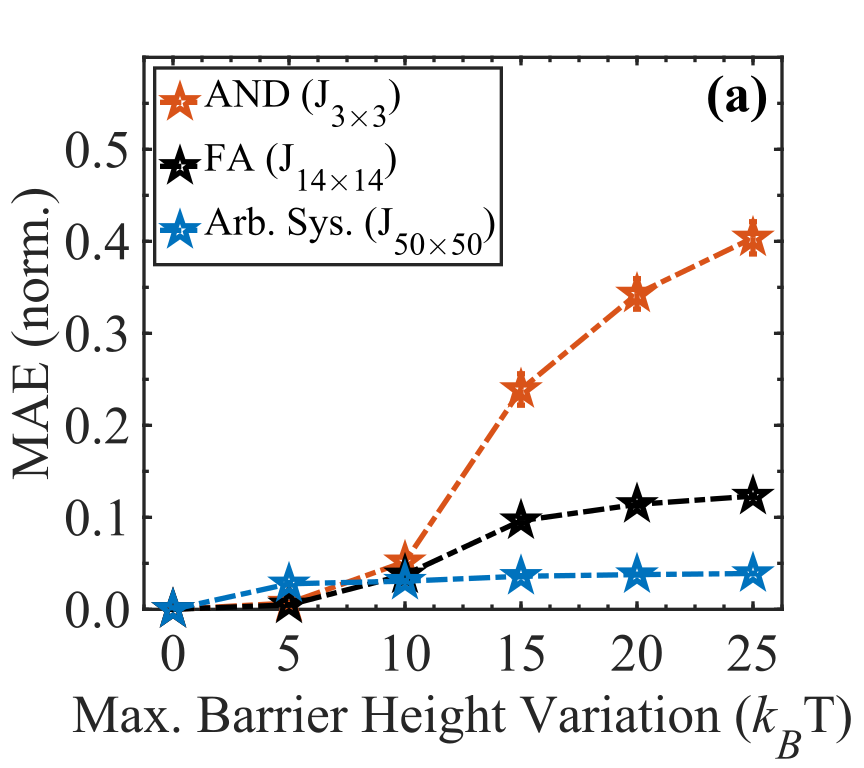}
    \includegraphics[width=0.49\linewidth]{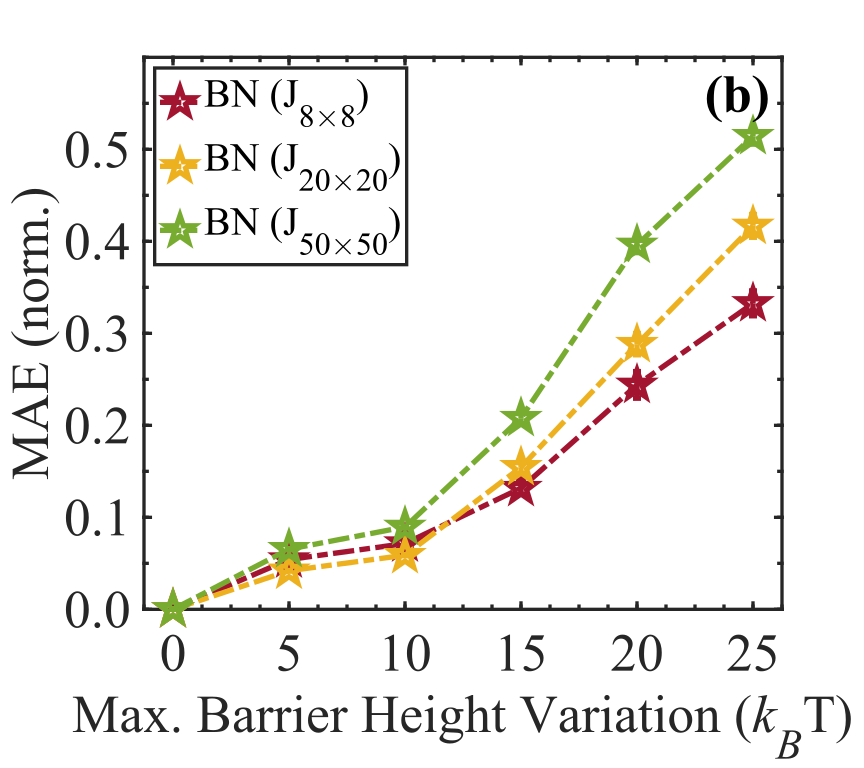}
    \caption{Normalized MAE from energy barrier variability for (a) EMOA and (b) PGA with different network sizes.} 
    \label{fig7}
\end{figure}

It is clear from (\ref{eq:magnet-state-retention}) that a small variation in the energy barrier $U$ can lead to a large variation in the expected state retention time $\tau$. This translates to a circuit encountering widely different time scales or a large dynamic range of operation within its individual components. This can lead to significant issues with the operational viability of a circuit built from p-bits. We, therefore, analyze the effect of energy barrier variation on the performance of the networks. As a result of the energy barrier variation, the magnetic states of different nanomagnets update at different times than the ideal case (assuming $0~k_BT$ energy barrier), leading to an overall error in the output quantity. Fig.~\ref{fig7} shows the normalized MAE for EMOA and PGA arising from energy barrier variability. We find that for both classes of problems, the error percentage is small (within $\sim 10\%$) up to an energy barrier of $\sim 10~k_BT$. For EMOA, the impact of a high energy barrier in a small network is severe ($\sim 40\%$ error), while the large network seems more forgiving in terms of error ($\sim 4\%$ error). On the contrary, the trend is the opposite in the case of PGA. We find a maximum of $\sim 50\%$ error for a large-sized BN. We note that we do not include the characteristics distortion variability while taking into account the energy barrier variability.
%%%%%%%%%%%%%%%%%%%%%%%%%%%%%%%%%%
%%%%%%%%%%%%%%%%%%%%%%%%%%%%%%%%%%
%%%%%%%%%%%%%%%%%%%%%%%%%%%%%%%%%%
\begin{figure}[!htbp]
    \centering
    \includegraphics[width=0.49\linewidth]{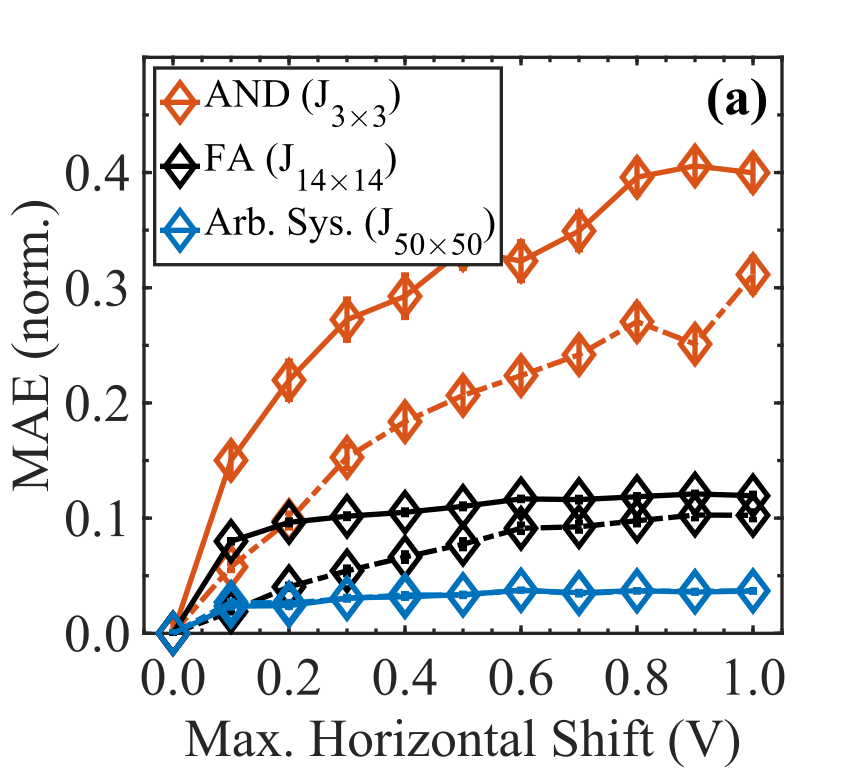}
    \includegraphics[width=0.49\linewidth]{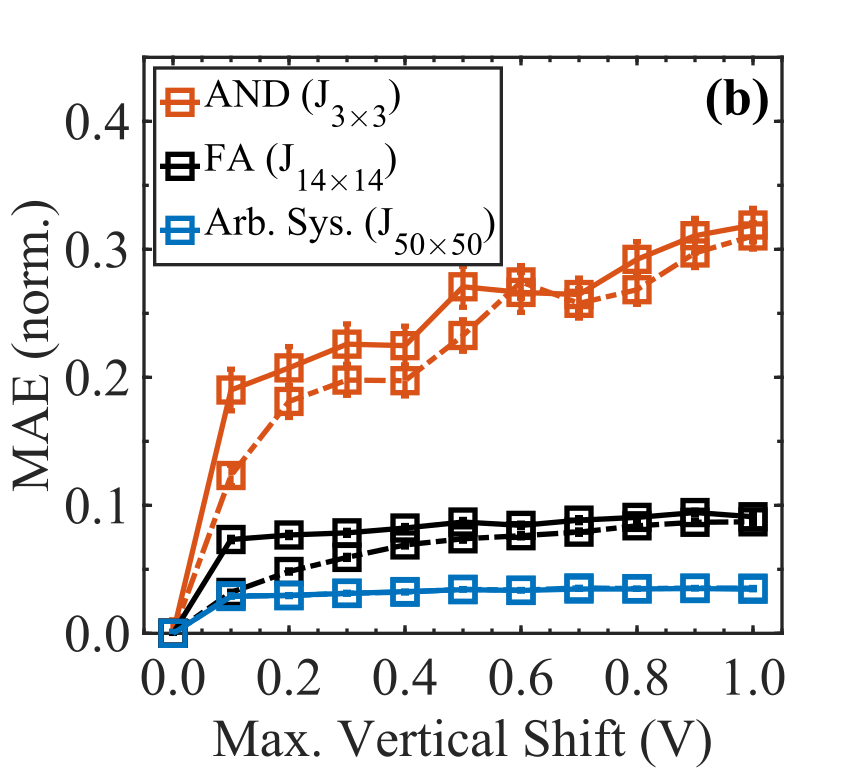}
    \includegraphics[width=0.49\linewidth]{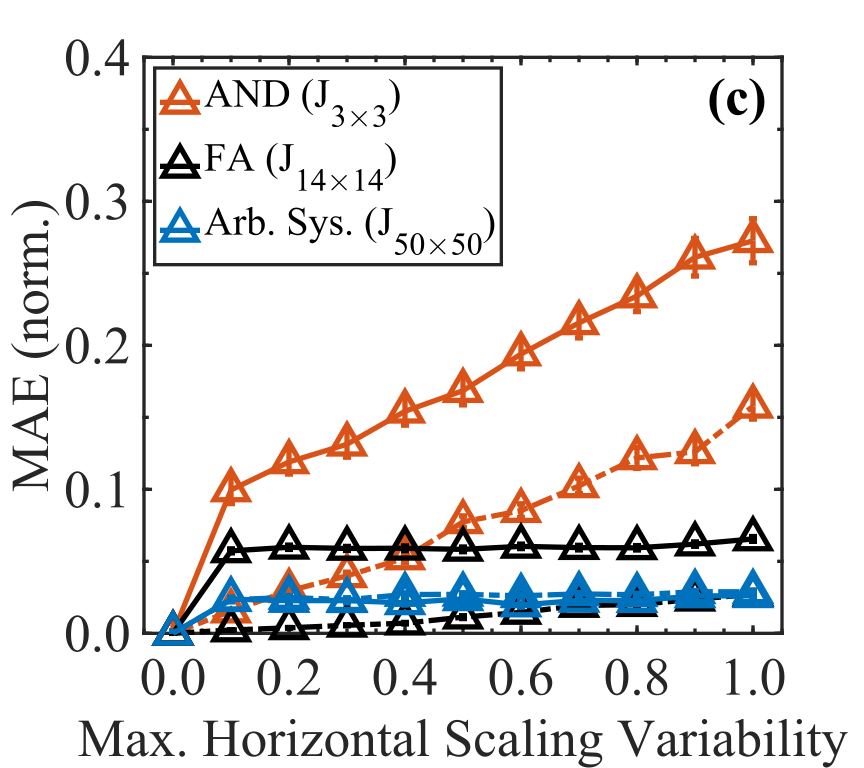}
    \includegraphics[width=0.49\linewidth]{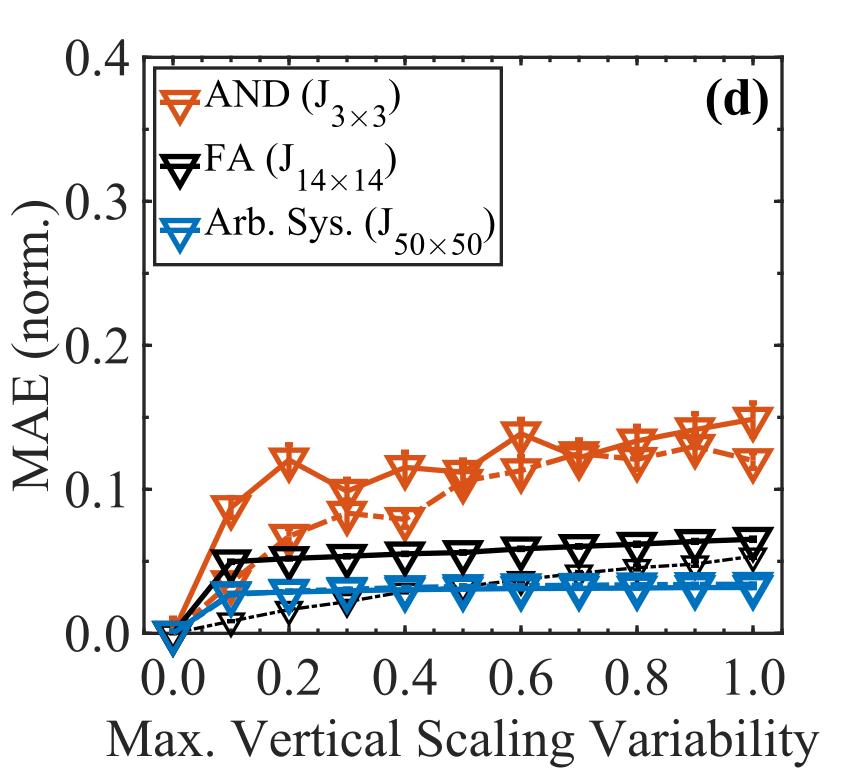}
    \caption{Normalized MAE calculated using sampling technique (dashed line) vs. simulated annealing technique (solid line) for EMOA for (a) horizontal shifting, (b) vertical shifting, (c) horizontal scaling, and (d) vertical scaling.} 
    \label{fig8}
\end{figure}
\subsection{Sampling vs. Simulated annealing}
The results discussed above for EMOA are calculated using the sampling technique, based on a fixed interaction strength $\kappa$ (pseudo-inverse temperature) throughout the simulation, and run the simulation long enough time ($10^6$ steps) so that the p-bits visit primarily the low-energy state. Fig.~\ref{fig8} shows the MAE using simulated annealing in comparison with the sampling technique. We vary $\kappa$ from $0.5$ to $5$ after every $2 \times 10^5$ steps while calculating the output using the simulated annealing technique. We find that the error percentage is slightly higher for all types of characteristics distortions for the simulated annealing technique. We conjecture that this is because the sampling method, when run long enough, can cover the system's phase space better ergodically than a linear simulated annealing schedule, which is in essence a guided importance sampling for a shorter time, may not be able to sample the phase space as comprehensively to discover the true ground state. This may be improved by more complex annealing schedules, which we do not discuss further.
\section{CONCLUSION}\label{conclusion}
In summary, we quantify the impact of non-idealities in computational networks built from LBM-based BSNs using two different techniques. In all the possible variances studied in this work, the error shows a sub-linear saturation at the extremal device variability points for EMOA, while in the PGA, the error grows linearly to super-linearly. We conjecture that this is because, in EMOA, the system tries to seek a single thermodynamically favorable fixed point in a finite phase space, which limits the growth of error, whereas, in PGA, there is no similar principle that can check the growth of the error. Additionally, running multiple samples of the same problem with different random seeds (thereby simulating the ``real world'') helps in reducing the variance of the error, but not its mean value. This suggests that for a certain amount of device variability, the average error is fixed, which may be estimated or characterized beforehand, and the results are certified accordingly. These findings may provide critical design insights for building suitable LBM-based hardware accelerators.

\section*{ACKNOWLEDGMENTS}\label{acknowledgments}
This work is supported in part by the NSF I/UCRC on Multi-functional Integrated System Technology (MIST) Center; IIP-1439644, IIP-1439680, IIP-1738752, IIP-1939009, IIP-1939050, and IIP-1939012. We thank Kerem Yunus Camsari and Faiyaz Elahi Mullick for useful discussions. All the calculations are done using the computational resources from High-Performance Computing systems at the University of Virginia (Rivanna).

\end{document}